\documentclass[aps,prl,twocolumn,superscriptaddress,floatfix,showpacs,reprint]{revtex4-1}
\usepackage{amsmath}
\usepackage{epsfig}
\usepackage{amssymb}
\usepackage{alltt}
\usepackage[latin1]{inputenc}
\usepackage[T1]{fontenc}
\usepackage[english]{babel}
\usepackage{accents}
\usepackage{hyperref}

\usepackage[dvipsnames]{xcolor}
\usepackage{upgreek}
\usepackage{bbold}
\usepackage{siunitx}
\usepackage{relsize}
\usepackage{pdfpages}

\usepackage[normalem]{ulem}

\newcommand{\tr}{\textrm{Tr}}
\newcommand{\lind}{\mathcal{L}}
\newcommand{\proj}{\mathcal{P}}
\newcommand{\hilb}{\mathcal{H}} 
\newcommand{\aver}[1]{\langle #1 \rangle}

\newcommand{\ii}{{\rm i}}
\newlength{\dhatheight}

\newcommand{\nohat}[1]{#1}
\def \nl{n_l}
\def \nr{n_r}

\def \lindstrong{\lind_0}
\def \lindpert{\mathcal{V}}
\def \lindeff{\tilde{\lind}^\lowsp}

\def \lowsp{{\Lambda_0}}
\def \bsp{\Omega_{B}}
\def \bspbis{\Omega_{b}}
\def \hspbare{\Lambda}
\def \hspalpha{{\Lambda_{\alpha}}}
\definecolor{redb}{RGB}{110,45,0}
\definecolor{greenb}{RGB}{45,110,0}

\def \element{\in}
\def \infinity{\infty}

\newcommand{\paragraphs}[1]{\paragraph{#1}}

\newcommand{\bonn}{University of Bonn, HISKP, Nussallee 14-16, 53115 Bonn, Germany}
\newcommand{\sutd}{Singapore University of Technology and Design, 20 Dover Drive, 138682 Singapore.}

\begin{document}

\title{Two-time Correlations Probing the Dynamics of Dissipative Many-Body Quantum Systems: Aging and Fast Relaxation} 

\author{Bruno Sciolla}
\affiliation{\bonn}

\author{Dario Poletti}
\affiliation{\sutd}

\author{Corinna Kollath}
\affiliation{\bonn}

\date{\today}

\begin{abstract}

Two-time correlations are a crucial tool to probe the dynamics of many-body systems.
We use these correlation functions to study the dynamics of dissipative quantum systems.
Extending the adiabatic elimination method, we show that the correlations can display two distinct behaviors, depending on the observable of interest:
a fast exponential decay, with a timescale of the order of the dissipative coupling, or a much slower dynamics. 
We apply this formalism to bosons in a double well subjected to phase noise. While the single-particle correlations decay exponentially, the density-density correlations display slow aging dynamics.
We also show that the two-time correlations of dissipatively engineered quantum states can evolve in a drastically different manner compared to their Hamiltonian counterparts. 
\end{abstract}

%05.70.Ln 	Nonequilibrium and irreversible thermodynamics     
%03.75.Kk 	Dynamic properties of condensates; collective and hydrodynamic excitations, superfluid flow 
%03.65.Yz 	Decoherence; open systems; quantum statistical methods 
%67.85.-d 	Ultracold gases, trapped gases
%67.85.De 	Dynamic properties of condensates; excitations, and superfluid flow 
%37.10.Vz 	Mechanical effects of light on atoms, molecules, and ions 

\pacs{03.65.Yz, 03.75.Kk, 67.85.-d, 05.70.Ln} 
\maketitle

\paragraphs{Introduction.} 
Not only are quantum systems characterized by their ground states, but also, very importantly, by their dynamical properties. 
Key aspects of these dynamic properties are captured by  {\it two-time}
correlation functions such as $\aver{A(t_2)B(t_1)}$.
Here $A$ and $B$ are observables, $t_1$ and $t_2$ two times and $\aver{\dots}={\rm tr}(\rho \dots)$ is the average over the density matrix of the system $\rho$. 
At equilibrium, the time-translation invariance leads to a dependence of the correlations on $t_2-t_1$ (or equivalently on one frequency only).
Many standard experimental probes such as ARPES, conductance measurements and neutron scattering in solids~\cite{AshcroftMermin1976}, or radio-frequency~\cite{StewartJin2008}, Raman or modulation spectroscopy~\cite{StoeferleEsslinger2004} in quantum gases rely on this relation.

An important example of two-time correlations from the area of quantum optics is the $g_2$-function,
\begin{align}
g_2(t_2,t_1)=\frac{\langle b^\dagger(t_1)  b^\dagger(t_2) b(t_2) b(t_1)\rangle}{\langle b^\dagger(t_1) b(t_1)\rangle \langle b^\dagger(t_2) b(t_2) \rangle}
\end{align}

which measures the second order coherence of a light field or of an atomic gas~\cite{ScullyZubairy1997}.
This observable shows bunching at small times for a thermal gas, and is constant for a (canonical) Bose-Einstein condensate, signaling second order coherence~\cite{YasudaShimizu1996,OettlEsslinger2005, GuarreraOtt2011}. Very recently, measurements of the $g_2$-correlation function have shown bunching of a continuously pumped photon Bose-Einstein condensate, manifesting the grand-canonical nature of the steady state \cite{SchmittWeitz2014}. 
This demonstrates the change of dynamic correlations in the presence of dissipation (see for example \cite{CarusottoCiuti2013, RitschEsslinger2013,HoeningFleischhauer2012,SiebererDiehl2013,LesanovskyGarrahan2013b,TaeuberDiehl2014}).

In contrast to thermal equilibrium and steady state situations, away from equilibrium the correlations acquire a dependence on both times $t_1$, $t_2$. This dependence is used in order to characterize, for example, the non-equilibrium physics of glasses and phase ordering kinetics \cite{Cugliandolo2002,RitortSollich2003,BerthierBiroli2011}.
These systems display {\it aging}, which is characterized by a slow, non-exponential, relaxation of correlations, by the breaking of time-translation invariance, and the presence of dynamical scaling, i.e. the correlations depend on the ratio $t_2/t_1$ only~\cite{HenkelSanctuary2007}.

In this article we develop a framework based on adiabatic elimination method \cite{ScullyZubairy1997,CarmichaelBook,BreuerPetruccione2002,GardinerZollerBook,Garcia-RipollCirac2009,ReiterSorensen2012,Kessler2012} to investigate the properties of two-time correlation functions in interacting many-body systems coupled dissipatively to a Markovian bath. Depending on the observables considered, the correlation functions fall into two different classes: one with a fast exponential decay, on the timescale set by the dissipation, and the other with a slow evolution which depends on the interplay between the dissipative and unitary dynamics. We apply our framework to strongly interacting bosonic atoms in a double well under the influence of phase noise. Remarkably, the density-density correlations follow an aging dynamics in the configuration space, characterized by a power-law dependence on the ratio of the two times, $\aver{ n_r(t_2)n_r(t_1)}/\aver{ n^2_r(t_1)} \propto \sqrt{t_1/t_2}$. Here $n_r$ denotes the number 
operator on the right well. In contrast, the single particle correlations decay 
exponentially fast in time.

Additionally, we point out that the dynamics of dissipatively engineered quantum states \cite{MuellerZoller2012,YiZoller2012, Daley2014} can be very different from the dynamics of their Hamiltonian counterparts. For example the normal and anomalous Green functions decay exponentially fast for a dissipatively engineered BCS state, while they oscillate under the BCS Hamiltonian evolution.

\begin{figure}
\includegraphics[width = 6.8cm]{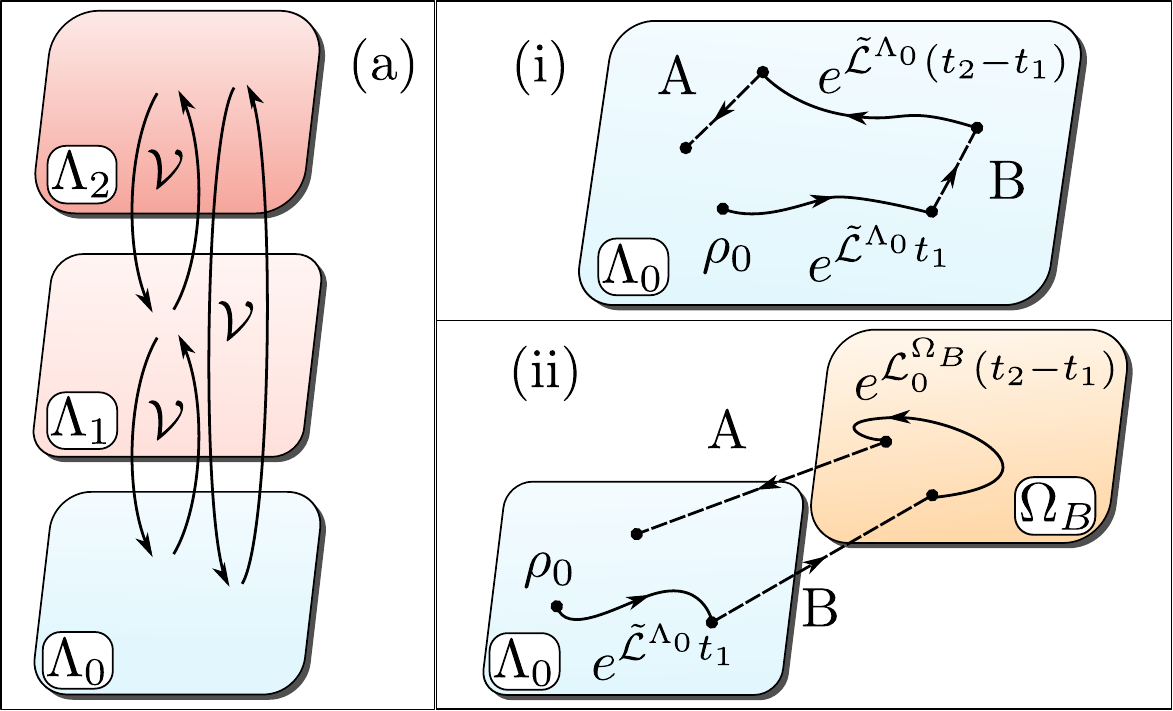}
\caption{\label{fig:subspaces} (color online) (a) Sketch of the decoherence free subspace, $\Lambda_0$, and two higher dissipative subspaces, $\Lambda_1, \Lambda_2$. (i) and (ii) Sketch of the different behavior of two-time correlations. (i) Case of an evolution within the decoherence free subspace $\lowsp$. (ii) Case of two-time correlations where the operator $B$ maps onto the subspace $\bsp$ and where the dynamics endowed by $\lind$ does not create transitions to $\lowsp$. Due to the evolution within $\bsp$, the correlations decay typically like $e^{-\lambda^R_{\bsp} (t_2-t_1)}$.}
\end{figure}

\paragraphs{Two-time correlations within the adiabatic elimination approach.}
We consider a quantum system evolving in contact with a Markovian bath. The dynamics of the density matrix $\rho$ of such a system is well described by the master equation
\begin{equation}
\label{eq:full_lind}
\partial_t \rho = \lind (\rho) = - \frac \ii {\hbar} [H,\rho] +\mathcal{D}(\rho) 
\end{equation}
consisting of a unitary part induced by the Hamiltonian $H$ and a dissipative term of Lindblad form
\begin{align}
\mathcal{D}(\rho)= \frac{\gamma}{2} \sum_m \left( 2 K_m \rho K_m^\dagger - K_m^\dagger K_m \rho - \rho K_m^\dagger K_m \right). \label{eq:diss0}
\end{align}
Here $K_m$ are quantum jump operators labeled by the index $m$ and $\gamma$ is the coupling strength to the bath. We assume that the Hamiltonian has two contributions $H=H_D+ H_\nu$, such that one can determine the eigenvalues and eigenstates of  $\lindstrong = -\left(\ii/\hbar\right) [H_D,\cdot] +\mathcal{D}$, i.e.~$\lindstrong \rho_\lambda = (-\lambda^R +  \ii \lambda^I)\rho_\lambda$ (where $\lambda^R \geq 0$ and $\lambda^I$ are real numbers). We consider the case in which $\lambda^R$ either vanishes or lies in bands [typically separated by gaps of the order $O(\gamma)$]. We call $\hspalpha$ the subspace of right eigenvectors with the same $\lambda^R_\alpha$. In particular, $\lowsp$ corresponds to $\lambda^R_0= 0$  and is called the decoherence free subspace. The second part $H_\nu$ instead is assumed to be weak and we define \mbox{$\lindpert (\rho) = -\ii/\hbar [H_\nu,\rho]$} [see Fig.\ref{fig:subspaces}(a)].

In such a situation, the density matrix elements belonging to a subspaces $\hspalpha$ with $\alpha\ne 0$
decay exponentially at a rate $ O(\gamma)$.
Thus, for times $t\gg \gamma^{-1}$, the dynamics is dominated by an effective dynamics in the decoherence free subspace $\lowsp$.
This effective dynamics can be obtained by adiabatically eliminating the faster decaying subspaces $\hspbare_{\alpha \ne 0}$ and is given by~\footnote{See Supplementary material for the details of this derivation.}
\begin{eqnarray}
 &&\frac{d}{dt}\rho^\lowsp = \lindeff \left(\rho^\lowsp \right) \label{eq:gen_adiab}\\
 &&\lindeff = \lindstrong^\lowsp - \sum_{\alpha\ne 0} \lindpert^{\lowsp \hspalpha} (\lindstrong^{\hspalpha})^{-1} \lindpert^{\hspalpha\lowsp}. \label{eq:SW2}
 \end{eqnarray}
Here we defined the projection of the density matrices into the subspace $X$ by $\rho^X = \proj^X (\rho) $ and the reduction of superoperators $\mathcal{O}$ onto $\mathcal{O}^{XY}: X\to Y$ as $\mathcal{O}^{YX} = \proj^Y \mathcal{O} \proj^X$, and $\mathcal{O}^{X} = \proj^X \mathcal{O} \proj^X$. We use the notation that a superoperator acts on everything to its right. 
The first term in Eq.~\ref{eq:SW2} represents the direct dynamics within $\lowsp$, whereas the second term represents the induced dynamics by virtual excitations to the $\hspalpha$ subspaces.

In the following we show that the adiabatic elimination formalism allows one to gain deep insight into the dynamics of two-time correlations
\begin{align}
& \langle A(t_2) B(t_1) \rangle = \tr \left[ A e^{\lind (t_2-t_1)}  \left( B  e^{\lind t_1} \nohat{\rho}_0  \right) \right]
\end{align}
for times $t_1,t_2-t_1\gg \gamma^{-1}$. Here $\nohat{\rho}_0$ is the initial density matrix.
% which we consider to be in $\Lambda_0$.
% This is indeed the typical scenario, since after a time of the order of $gamma^{-1}$ the contributions to the state in higher subspaces have decayed and the state lies dominantly within $Lambda_0$. 
The computation of such correlations requires several steps: the time evolution of the density matrix for a time $t_1$, the action of $B$, the evolution of the resulting operator for time $t_2-t_1$, the action of the operator $A$ and taking the trace. 
For a time $t_1$ much larger than $\gamma^{-1}$, the time-evolved operator $e^{\lind t_1} \nohat{\rho}_0$ can be well approximated by $e^{\lindeff t_1} \nohat{\rho}_0$.
The subsequent action of operator $B$ has different consequences, depending on whether the subspace $\bsp$ reached by the application of the operator $B$ and the following time evolution has an overlap with the decoherence free subspace $\lowsp$ or not.
In the following we will concentrate on two cases: (i) $\bsp$ is a subspace of $\lowsp$ [Fig.\ref{fig:subspaces}(i)], (ii) $\bsp$ has no overlap with $\lowsp$ nor with any subspace of $\hspalpha$ connected to $\lowsp$ by repeated action of $\lindpert$ [Fig.\ref{fig:subspaces}(ii)] \footnote{For example, the space $\bsp$ belongs to a different symmetry sector (here related to the atom number) than all the $\hspalpha$.}.        

In the first case (i), the dominant contribution is the effective evolution in $\lowsp$:
\begin{align}
& \langle A(t_2) B(t_1) \rangle \approx \tr \left[ A e^{\lindeff (t_2-t_1)}   \left(B  e^{\lindeff t_1} \nohat{\rho}_{0}  \right)  \right] \label{eq:lowerspobs}
\end{align}
In contrast, in the second case (ii), the evolution between the action of the operator $B$ at time $t_1$ and operator $A$ at time $t_2$ takes place mostly in $\lindstrong^{\bsp}$ thus giving:   
\begin{align}
& \langle A(t_2) B(t_1) \rangle \approx \tr \left[  A e^{\lindstrong^{\bsp} (t_2-t_1)} \left( B  e^{\lindeff t_1} \nohat{\rho}_0 \right) \right] \label{eq:higherspobs}
\end{align}

In the first case, the correlations can be long-lived and can lead to a slow nontrivial dynamics within $\lowsp$, as we shall see in an example. In the second case the correlations typically decay exponentially fast like $e^{-\lambda_{\bsp}^R(t_2-t_1)}$~\footnote{Note that the imaginary part of $\lambda_{\bsp}$ may also induce oscillations.}. 
The two cases \eqref{eq:lowerspobs} and \eqref{eq:higherspobs} are two extreme examples of the slow and fast dynamics of the two-time correlations that may occur depending on the observable of interest and on the properties of the Lindblad evolution.
Intermediate situations can also be dealt with within the developed formalism.

\paragraphs{Two-sites Bose-Hubbard model.}      
In the following we investigate the two-time correlation dynamics in a two-sites Bose-Hubbard model in the presence of phase noise \cite{LouisSavage2001,BrennenWilliams2005,HuangMoore2006,KhodorkovskyVardi2008,GerbierCastin2010,FerriniHekking2010,Bar-GillKurizki2011,WitthautWimberger2011,PolettiKollath2012,BuchholdDiehl2014,MendozaArenasClark2013, PolettiKollath2013,Daley2014}. The Hamiltonian is given by $H_\nu = -J (b^\dagger_l b_r + b^\dagger_r b_l)$ and $H_D= U/2 \sum_{i=l,r} n_i(n_i-1)$ where $b_i^{(\dagger)}$ are bosonic annihilation (creation) operators on the left (right) site $i=l\;(r)$. The dephasing noise is represented by local quantum jump operators $K_i = n_i$.
It has been previously shown that the equal time quantities display interesting power-law dynamics due to anomalous diffusion in the configuration space~\cite{PolettiKollath2012}. 

{\it (i) Density-density correlations.}
We show in the following that the evolution of the connected two-time density-density correlations $\langle \nohat{n}_r(t_2) \nohat{n}_r(t_1) \rangle_c=\langle \nohat{n}_r(t_2) \nohat{n}_r(t_1) \rangle-\langle \nohat{n}_r(t_2)\rangle\langle \nohat{n}_r(t_1) \rangle$, where we take the right site $i=r$ for definiteness, is a typical example of a slow evolution determined by Eq.~(\ref{eq:lowerspobs}) restricted to the decoherence free subspace. Indeed, the space $\Omega_{n_r}$, created by the application of the operator $n_r$ on an element of decoherence free subspace $\lowsp$, lies within it. Here $\lowsp$ consists of all diagonal density matrices in the Fock space $\rho^{\lowsp} = \sum_{\nl} \rho_{\nl}^{\lowsp} |\nl, \nr  \rangle \langle \nl, \nr |$. The state $|\nl, \nr  \rangle $ has $n_{l(r)}$ atoms on the left (right) site with the relation $\nr = N-\nl$, and $N$ is the total particle number.

\begin{figure}
\includegraphics{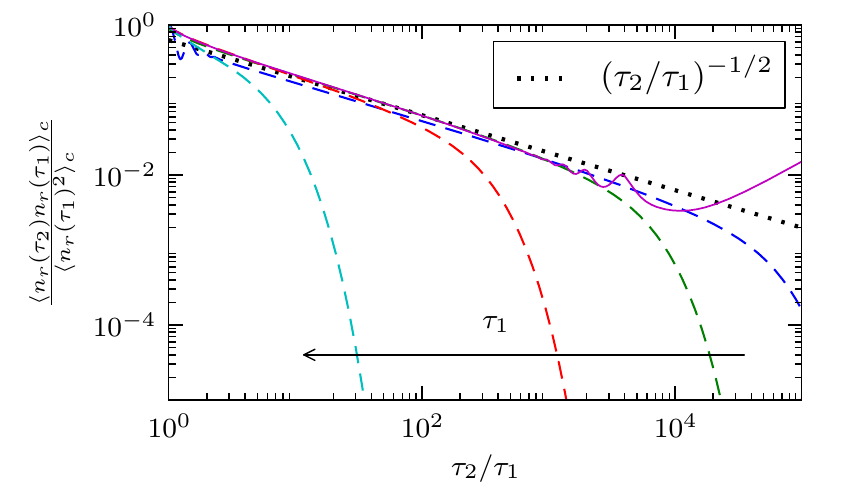}
\caption{\label{fig:autocor-t-rescaled-loglog}(color online) (a) Rescaled density-density correlations versus the ratio of the rescaled times $\tau_2/\tau_1$ from the master equation (\ref{eq:full_lind}), for $U/J=20$, $\hbar \gamma/J=5$, $N=50$. Dashed lines correspond to a fixed value of 
$\tau_1 =  10^{-6}, 10^{-5}, 2 \times 10^{-4}, 10^{-2}$
from right to left. The solid line corresponds to a fixed value of $\tau_2= 10^{-3}$.  The correlations follow the scaling solution $\propto \sqrt{\tau_1/\tau_2}$ (thick dotted line) for intermediate times.
}
\end{figure}

The resulting slow dynamics of the connected correlations $\langle \nohat{n}_r(t_2) \nohat{n}_r(t_1) \rangle_c$ exhibits aging at intermediate times (Fig.~\ref{fig:autocor-t-rescaled-loglog}). This can be seen both in the numerical results obtained by solving the master equation~\eqref{eq:full_lind} and approximate analytical results obtained from the effective evolution~\eqref{eq:lowerspobs} which we will derive in the following. These analytical results give a power law decay of the rescaled correlations with the ratio of the two times $t_2/t_1$ of the form
\begin{align}
\frac{ \langle \nohat{n}_r(t_2) \nohat{n}_r(t_1) \rangle_c}{\langle \nr(t_1)^2 \rangle} = \frac {|C|} { \sqrt{2} } \frac { \Gamma(5/4) \Gamma(1/4) } { \Gamma(3/4) } \sqrt{\frac{t_1}{t_2}}. \label{eq:scaling_nn_result}
\end{align} 
with a real constant $C$. Here $\Gamma$ denotes the $\Gamma$-function~\cite{Abramowitz}.

In the following we derive {\it analytically} this scaling form of the two-time correlation function using the adiabatic elimination for large $N$.
For this purpose, we introduce
the continuous variable $x = n_l/N-1/2 \element [-1/2,1/2]$ which measures the occupation of the left well and we define the distribution $p(x,\tau_1) = N\rho^{\lowsp}_{n_l}(t_1=\tau_1 \tilde{t}\;)$ with the rescaled time $\tau_1=t_1/\tilde{t}$ (and $\tau_2=t_2/\tilde{t}\;$) and $\tilde{t}={2 U^2 N^2/}{J^2 \gamma}$.
At large interactions, Eq.~\eqref{eq:gen_adiab} can be written as~\footnote{See~\cite{PolettiKollath2012} and Eq. (7S) in the Supplementary material.}
\begin{align}
&\partial_\tau p(x,\tau) = \partial_x\left[ D(x)\partial_x p(x,\tau)\right] \label{eq:cont_p}
\end{align}
with the configuration space dependent diffusion function $D(x) = 1/(4x^2) -1$. Starting from the initially balanced configuration $p(x,0)=\delta(x)$, the diverging part of the diffusion function $D(x)\approx 1/(4x^2)$ causes a fast initial dynamics characterized by an anomalous diffusion~\cite{PolettiKollath2012}. This anomalous diffusion signals the scaling regime and the distribution $p$ takes the form $p(x,\tau_1^-) = \frac{\sqrt{2}}{\Gamma(1/4)}\tau_1^{-1/4} \exp [-x^4/(4 \tau_1)] $.
At longer times (for $\tau_1 \gtrsim 10^{-2}$ for the parameters shown in Fig.~\ref{fig:autocor-t-rescaled-loglog}), the function $p(x,\tau_1^-)$ reaches the boundaries $x=\pm1/2$ and leaves the scaling regime. It then approaches the stationary infinite temperature state $p(x,\infinity)=1$ exponentially, due to the finite particle number $N$.

To compute $\langle \nohat{n}_r(\tau_2) \nohat{n}_r(\tau_1) \rangle$, we multiply $p(x,\tau_1^-)$ by $n_r = N(1/2 - x)$.
Due to the linearity of the evolution equation, the new distribution can be evolved separately for the symmetric $p^S(x,\tau_1^+)=(N/2) p(x,\tau_1^-)$ and the antisymmetric $p^A(x,\tau_1^+)=-N x \,p(x,\tau_1^-)$ part. 
The symmetric part has a constant contribution $\int dx\; N(1/2-x) p^S(x,\tau_2) = N^2/4$ to the correlations. The antisymmetric part \mbox{$p^A(x,\tau_1^+)= \frac{-N \sqrt{2}}{\Gamma(1/4)}\frac{x}{\tau_1^{1/4}} \exp [-x^4/(4\tau_1)]$} has a non-trivial contribution and we will focus on that.
Note that $p^S$ and $p^A$ are not normalized probability distributions. 

\begin{figure}
\includegraphics{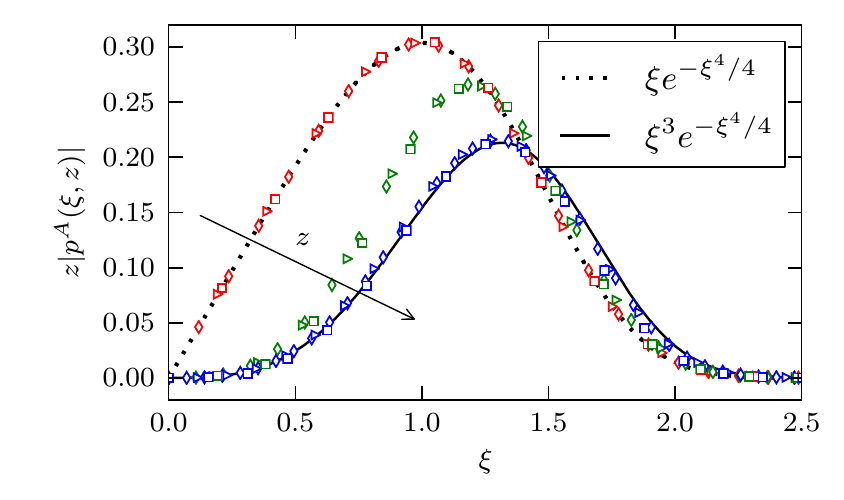}
\caption{\label{fig:g_scaling_b}
(color online)  Antisymmetric part of the density matrix elements $|p^A(\xi,z)|$ rescaled by $z={\tau_2}/{\tau_1}$, versus $\xi={x}/{\tau_2^{1/4}}$, for several scaling times $z = 1.03,1.5,8$ ($z$ arrow) and for different values of 
$\tau_1 = 2.7\times 10^{-8}$ (triangles), $2\times 10^{-7}$
(diamonds) from the master equation (\ref{eq:full_lind}) with $U/J = 20$, $\hbar \gamma/J=5$ and $N=50$.
The square symbols are obtained using Eq.~\eqref{eq:gen_adiab} for $N=400$.
The dots and the line represent, respectively, the scaling initial conditions $\xi e^{-\xi^4/4}$ and separable solution $\xi^3 e^{-\xi^4/4}$.
}
\end{figure}

The following evolution of $p^A(x,\tau_2;\tau_1)$ as a function of $\tau_2$ is determined by Eq.~\eqref{eq:cont_p}, with initial condition \mbox{$p^A(x,\tau_2=\tau_1;\tau_1)=p^A(x,\tau_1^+)$}.
We now recast Eq.~\eqref{eq:cont_p} into a scale invariant form using the scaling variables $\xi = x/\tau_2^{1/4}$ and $z =  \tau_2/\tau_1$. For sufficiently small values of $\xi$ we obtain the scaling equation:
\begin{align}
\xi \partial^2_\xi p(\xi,z) +(\xi^4 -2) \partial_\xi p(\xi,z) -4 \xi^3 z \partial_z p(\xi,z) = 0, \label{eq:scal}
\end{align}
where one sees that all dependency on $x$, $\tau_1$ and $\tau_2$ can be absorbed in $\xi$ and $z$.
This equation has a separable asymmetric solution:
\begin{align}
p^A(\xi,z) = C \frac{\xi^3}{z} \exp(-\xi^4/4) \label{eq:sca_sol}
\end{align}
where $C<0$ is the same constant as in Eq.~\eqref{eq:scaling_nn_result}. 
In Fig.~\ref{fig:g_scaling_b} we show the evolution of the antisymmetric part of the density matrix obtained from the full Markovian evolution of Eq.~\eqref{eq:full_lind} and from the restriction to the lower subspace Eq.~\eqref{eq:gen_adiab}, as a function of the scaling variables.
We find an excellent collapse for each chosen value of $z$, for various $\tau_1$, which confirms the validity of the scaling equation~\eqref{eq:scal}. At short times $\tau_2 \sim \tau_1$, the distribution follows the initial distribution $p^A(\xi,z=1)\propto \xi e^{-\xi^4/4}$ and converges for larger $z$ towards the separable solution \eqref{eq:sca_sol}.
The correlations are finally obtained applying the operator $\nohat{n}_r$ and taking the trace.
Gathering the result with the symmetric contribution and recalling that the local fluctuations grow with time like $\langle \nr(\tau_1)^2 \rangle = 2\Gamma(3/4)/\Gamma(1/4) \tau_1^{1/2}$ (Eq.(6) in \cite{PolettiKollath2012}), one obtains the result of Eq.~\eqref{eq:scaling_nn_result}.
Thus we recover the
existence of an algebraic decay as a function of $\tau_2/\tau_1$ [see Fig.~\ref{fig:autocor-t-rescaled-loglog}], which implies aging.

{\it (ii) Two-time single particle correlation function.} 
The two-time single particle correlation function $\langle\nohat{b}^\dagger_r(t_2) \nohat{b}_r(t_1) \rangle$, i.e.~the bosonic phase coherence, behaves very differently from the two-time density-density correlations.
The origin of this marked difference lies in the fact that the operator $\nohat{b}_r$, applied at time $t_1$, brings the density matrix to a space, $\Omega_{b}$, which has different atom numbers from the decoherence free subspace $\lowsp$ and thus follows Eq.~\eqref{eq:higherspobs} [cf.~Fig.~\ref{fig:subspaces}(ii)].

As before, the evolution up to time $t_1$ of $\rho^{\lowsp}$ follows Eq.~\eqref{eq:gen_adiab}.
The density matrix just after the application of $b_r$ at time $t_1$ can be represented by $\rho^{\Omega_{b}}(t_1^+) = \sum_{\nl}\rho_{\nl}^{\Omega_{b}}|\nl,N-1-\nl \rangle \langle \nl,N-\nl | $.
This operator $\rho^{\Omega_{b}}$ acts on the space $ \hilb(N)$ and leads to the space $\hilb(N-1)$, where $\hilb(N)$ is the Hilbert space with $N$ particles.
Since the evolution preserves the particle number between time $t_1^+$ and $t_2^-$, the operator remains in this subspace $\bspbis$ until the operator $b_r^\dagger$ is applied at time $t_2$. The slowest decaying eigenvalues of $\lindstrong$ in this operator space are given by $-{\gamma}/{2} + \ii (U/\hbar) (N-\nl -1)$ with $\nl\in [0,N]$. Thus using Eq.~\eqref{eq:higherspobs} we find that the evolution, to zero-th order in $J$, is given by \mbox{$\langle\nohat{b}^\dagger_r(t_2) \nohat{b}_r(t_1) \rangle = e^{- \frac{\gamma}{2} (t_2-t_1)} \sum_{n_l} n_{l} e^{\ii \frac U {\hbar} (N-n_l -1) (t_2-t_1)} \rho^{\lowsp}_{n_l}(t_1^-) $}.

\begin{figure}
\includegraphics{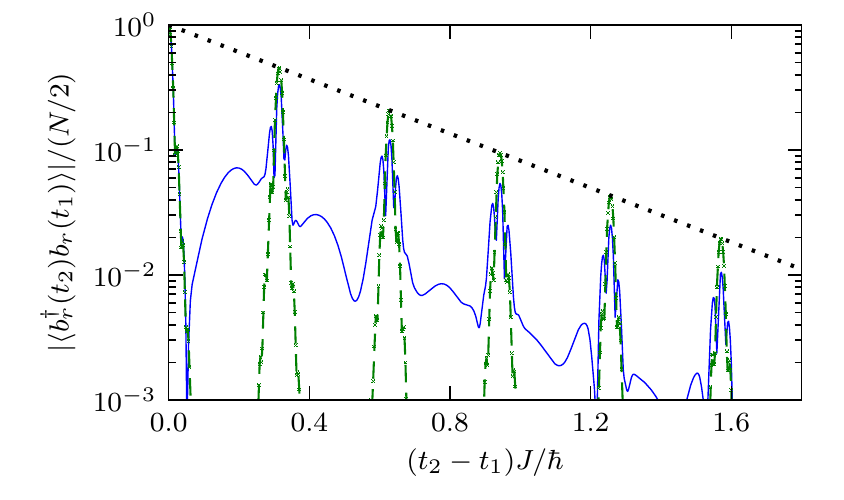}
\caption{\label{fig:bdagb_comp} 
(color online) Single particle correlations as a function of $(t_2-t_1)J/\hbar$ for $t_1 =256 J/\hbar$ (corresponding to $\tau_1= 10^{-3}$) and $U/J=20$, $\hbar \gamma/J=5$, $N =40$, from the master equation (solid line) and the zero-th order expansion in $J$ (dashed line).
The dotted line represents a pure exponential decay with decay constant $\gamma/2$.}
\end{figure}

The single particle correlations are depicted in Fig. \ref{fig:bdagb_comp}, where we compare this analytical solution to the master equation \eqref{eq:full_lind}, which includes tunneling within the $\bspbis$ subspace \footnote{The dynamics within the $\bspbis$ subspace is described in the supplementary material.}. As anticipated by the analytical expression, the single particle correlations decay approximately exponentially as $\langle\nohat{b}^\dagger_r(t_2) \nohat{b}_r(t_1) \rangle \sim e^{-\gamma(t_2-t_1)/2}$ and show coherent oscillations of periodicity $2 \pi \hbar/U$ caused by the interactions. Only the detailed structure of the exact results is not covered by the analytical approximation. The precise shape of the oscillations depends on the initial times $t_1$ through the initial distribution $\rho^{\lowsp}(t_1^-)$, whereas the characteristic decay and oscillations are independent of the value of $t_1$.

\paragraphs{Conclusion} 
We have developed a framework to determine and classify the dynamics of two-time correlation functions for dissipative many-body systems based on adiabatic elimination. 
We applied the developed approach to bosonic gases in a double well subjected to phase noise. Remarkably, in this situation, the density-density correlations exhibit aging, whereas single-particle correlations decay exponentially fast in time. This example shows that the interplay of the dissipative coupling and the unitary evolution can lead to a very rich dynamics which can be uncovered with our approach. 

Additionally, our results have far reaching consequences for the dissipative engineering of interesting quantum states \cite{MuellerZoller2012}. Indeed, within the formalism developed, one can readily show that the fascinating dynamic properties of a unitary state are not necessarily reproduced in their dissipatively engineered counterparts. 
For example, under the evolution due to the BCS mean-field Hamiltonian, the two-time single particle and anomalous Green functions, respectively~ $G^>_{\sigma}(\vec{k},t_2,t_1) = \tr \left[c_{\vec{k},\sigma}(t_2) c^\dagger_{\vec{k},\sigma}(t_1)\right]$ and $G^A(\vec{k},t_2,t_1) = \tr \left[ c_{\vec{k},\uparrow}(t_2) c_{-\vec{k},\downarrow}(t_1) \right]$, exhibit an oscillatory behavior when computed for the BCS ground state [here $c_{\vec{k},\sigma}\;(c^{\dagger}_{\vec{k},\sigma})$ annihilates (creates) a fermion with momentum $\vec{k}$ and spin $\sigma$].
In contrast, if for instance the BCS state is dissipatively prepared as the dark state of jump operators which are the BCS Bogoliubov excitation operators $K_{\vec{k},\sigma}=\gamma_{\vec{k},\sigma}$ with a coupling strength $\kappa$, both Green functions decay exponentially fast in time, i.e. $G^{>}_{\sigma}(\vec{k},t_2,t_1),\;G^{A}(\vec{k},t_2,t_1) \propto e^{-\kappa (t_2-t_1)/2}$, as a result of Eq.~\eqref{eq:higherspobs}~\footnote{The Green functions are computed in the supplementary material.}. Thus, if the goal is to access dynamic properties of a state, it is not sufficient to dissipatively engineer it, but, as discussed in~\cite{YiZoller2012} for the d-wave BCS state, an adiabatic passage to the unitary system has to be performed.

We acknowledge support from the SUTD start-up grant (SRG-EPD-2012-045), DFG and BCGS. We acknowledge fruitful discussions with M. Fleischhauer, S. Diehl and G. Kocher.

\bibliography{ref.bib}

\newpage


\section*{Supplementary material}

# Two-time Correlations Probing the Dynamics of Dissipative Many-Body Quantum Systems: Aging and Fast Relaxation
# Supplementary Material

Bruno Sciolla,[1] Dario Poletti,[2] and Corinna Kollath[1]

[1]*University of Bonn, HISKP, Nussallee 14-16, 53115 Bonn, Germany*
[2]*Singapore University of Technology and Design, 20 Dover Drive, 138682 Singapore.*

(Dated: July 17, 2014)

## DERIVATION OF THE EFFECTIVE EQUATION OF MOTION

In this section, we derive Eq.(4) and (5) in the main text, describing the effective dynamics of the density matrix in the decoherence free subspace. This equation is obtained using adiabatic elimination [1–6]. We recall the definition of the right eigenvectors of $\mathcal{L}_0$, satisfying $\mathcal{L}_0 \rho_{\lambda_i} = (-\lambda_i^R + i\lambda_i^I)\rho_{\lambda_i}$. We decompose an arbitrary density matrix $\rho^{\Lambda_0}$ belonging to $\Lambda_0$ into the right eigenvectors $\rho_{\lambda_{0,i}}$ of $\mathcal{L}_0$, $\rho^{\Lambda_0} = \sum_i \rho_i^{\Lambda_0} \rho_{\lambda_{0,i}}$. The eigenvectors $\rho_{\lambda_{0,i}}$ correspond to the eigenvalues $\lambda_{0,i} = -\lambda^R + i\lambda_i^I$ with $\lambda^R = 0$. The label $i$ allows to distinguish the different eigenvectors, which can have different imaginary eigenvalues. The coefficients of the density matrix can be represented by $\rho_i^{\Lambda_0} = \text{Tr}(\rho^{\lambda_{0,i}\dagger}\rho^{\Lambda_0})$ using the left eigenvectors $\rho^{\lambda_{0,i}}$, which satisfy $\rho^{\lambda_{0,i}}\mathcal{L}_0 = \lambda_{0,i}\rho^{\lambda_{0,i}}$. We use a similar decomposition on all the other subspaces $\Lambda_\alpha$. This allows us to write the evolution equations in terms of the elements $\rho_i^{\Lambda_0}$, $\rho_j^{\Lambda_\alpha}$.

With these notations, the evolution of the density matrix is fully described by:

$$\frac{d}{dt}\rho_i^{\Lambda_0}(t) = \lambda_{0,i}\rho_i^{\Lambda_0}(t) + \sum_{j,\alpha} \mathcal{V}_{ij}^{\Lambda_0\Lambda_\alpha}\rho_j^{\Lambda_\alpha}(t) \quad (1S)$$

$$\frac{d}{dt}\rho_j^{\Lambda_\alpha}(t) = \lambda_{\alpha,j}\rho_j^{\Lambda_\alpha}(t) + \sum_{k,\beta\neq\alpha} \mathcal{V}_{jk}^{\Lambda_\alpha\Lambda_\beta}\rho_k^{\Lambda_\beta}(t) \quad (2S)$$

In these equations we have used the operator elements $\mathcal{O}_{i,i'}^{\Lambda_\alpha\Lambda_\beta} = \text{Tr}\left[(\rho^{\lambda_{\alpha,i}})^\dagger \mathcal{O}\rho_{\lambda_{\beta,i'}}\right]$.

We can formally integrate the dynamics of the $\rho_j^{\Lambda_\alpha}$ elements for $\alpha > 0$ as

$$\rho_j^{\Lambda_\alpha}(t) = e^{\lambda_{\alpha,j}t}\rho_j^{\Lambda_\alpha}(0)$$
$$+ \sum_{k,\beta\neq\alpha} e^{\lambda_{\alpha,j}t}\int_0^t dt' e^{-\lambda_{\alpha,j}t'}\mathcal{V}_{jk}^{\Lambda_\alpha\Lambda_\beta}\rho_k^{\Lambda_\beta}(t') \quad (3S)$$

Here the first term corresponds to an exponential decay of the elements $\rho_j^{\Lambda_\alpha}(0)$. Assuming times of the order of $\gamma t \gtrsim 1$, we neglect these transient terms in the following. The second term has to be dealt with more carefully. It can be simplified by performing an integration by parts and neglecting higher order terms in $1/\lambda_{\alpha,j}$ for $\alpha > 0$.

Doing this we obtain

$$\rho_j^{\Lambda_\alpha}(t) \approx$$
$$\sum_i \frac{1}{\lambda_{\alpha,j}}\left[e^{\lambda_{\alpha,j}t}\mathcal{V}_{ji}^{\Lambda_\alpha\Lambda_0}\rho_i^{\Lambda_0}(0) - \mathcal{V}_{ji}^{\Lambda_\alpha\Lambda_0}\rho_i^{\Lambda_0}(t)\right]$$
$$+ O(1/\lambda_{\alpha,j}^2)$$

We can further neglect the remaining transient terms decaying like $e^{-\lambda_{\alpha,j}^R t}$ and we obtain

$$\rho_j^{\Lambda_\alpha}(t) \approx -\sum_i \frac{1}{\lambda_{\alpha,j}}\mathcal{V}_{ji}^{\Lambda_\alpha\Lambda_0}\rho_i^{\Lambda_0}(t) \quad (4S)$$

We can now derive the effective evolution of $\rho_i^{\Lambda_0}(t)$, by substituting the expression for the evolution of the elements $\rho_j^{\Lambda_\alpha}(t)$ in the higher subspaces into equation (1S),

$$\frac{d}{dt}\rho_i^{\Lambda_0}(t) = \lambda_{0,i}\rho_i^{\Lambda_0}(t) - \sum_{i',\alpha,j}\frac{1}{\lambda_{\alpha,j}}\mathcal{V}_{ij}^{\Lambda_0\Lambda_\alpha}\mathcal{V}_{ji'}^{\Lambda_\alpha\Lambda_0}\rho_{i'}^{\Lambda_0}(t)$$
$$(5S)$$

which can be written in operator form as:

$$\frac{d}{dt}\rho^{\Lambda_0}(t) = \mathcal{L}_0^{\Lambda_0}\rho^{\Lambda_0}(t) - \sum_\alpha \left[\mathcal{V}^{\Lambda_0\Lambda_\alpha}(\mathcal{L}_0^{\Lambda_\alpha})^{-1}\mathcal{V}^{\Lambda_\alpha\Lambda_0}\rho^{\Lambda_0}(t)\right]$$
$$(6S)$$

Thus, we have recovered equations (4) and (5) of the main text, which describe the effective evolution in the lowest subspace $\Lambda_0$.

## DISCRETE EFFECTIVE MASTER EQUATION FOR THE TWO-SITE BOSE-HUBBARD MODEL

A density matrix in the decoherence free subspace of the phase noise in the two-site Bose-Hubbard model can be written as $\rho^{\Lambda_0} = \sum_{n_l} \rho_{n_l}^{\Lambda_0}|n_l,n_r\rangle\langle n_l,n_r|$, where the particle number on the right site is fully determined by $n_r = N - n_l$. In the following we will use $n$ in order to label the left particle number. The effective equation of evolution within the decoherence free subspace [corresponding to Eq. (4) in the main text] reduces to coupled equations for the coefficients $\rho_n^{\Lambda_0}$. These are given by [7]:

$$\partial_t \rho_n^{\Lambda_0} = \frac{N^2}{\tilde{t}}\left[W_n(\rho_{n+1}^{\Lambda_0} - \rho_n^{\Lambda_0}) + W_{n-1}(\rho_{n-1}^{\Lambda_0} - \rho_n^{\Lambda_0})\right]$$
$$(7S)$$

where $W_n = \frac{(n+1)(N-n)}{(n-N/2+1/2)^2}$ and where $\tilde{t} = \frac{2U^2 N^2}{J^2 \gamma}$ is the relevant time scale of the evolution [8]. Taking the continuum limit $x = n/N - 1/2$ and $p(x,\tau) = N\rho_n^{\Lambda_0}(t = \tau\tilde{t})$, one obtains Eq. (10) in the main text.

## RANGE OF VALIDITY OF THE AGING REGIME

The range of validity of the scaling regime is approximately $1/\gamma\tilde{t} \ll \tau_1, \tau_2 \ll 10^{-2}$ for the parameters in Fig. 2 in the main text. In particular, the earlier $\tau_1$ is, the longer the correlations follow the scaling as a function of $\tau_2/\tau_1$ before deviations enter as $\tau_2 \gtrsim 10^{-2}$. For $\tau_1$ too small, like $\tau_1 = 10^{-6}$ in Fig. 2 in the main text, the scaling regime is not yet reached, and one can see oscillations due to the coherent dynamics for small $\tau_2/\tau_1$. We complement these results with additional computations for various different sets of the interaction strength, the dissipative coupling strength and the number of particles in Fig. 1S. The behaviour of all these curves is similar, showing a scale-invariant regime. We would like to stress that this is true even for low values of the interaction strength and the dissipative coupling, a regime in which we would not expect the adiabatic elimination to work well.

In the correlations for a fixed value of $\tau_2 = 10^{-3}$ (Fig. 2 in the main text), similar oscillations caused by the unitary dynamics can be observed at large $\tau_2/\tau_1$, i.e. small values of $\tau_1$. The lower bound $\gamma\tilde{t} \ll \tau_1, \tau_2$ is the condition of validity of the effective evolution in the decoherence free subspace (4). The upper bound coincides with the breakdown of the scaling regime for the equal time correlations [7] and it is caused by the finite number of particles. At large times $\tau_1$, the system approaches the stationary state. In this regime, the correlations decay exponentially with the time difference $\tau_2 - \tau_1$ like $\langle n_r(\tau_2) n_r(\tau_1)\rangle_c \sim e^{-(\tau_2-\tau_1)/T_\infty}$ with fixed decay time $T_\infty \approx 3.10 \times 10^{-3}$ (in rescaled time units).

## EVOLUTION OF THE SINGLE-PARTICLE CORRELATIONS

In Fig. 2S we show the dependency of the evolution of the single-particle correlations on the various parameters. One can readily see that the typical decay rate is $\gamma/2$, which is illustrated with the value $\hbar\gamma/J = 5$ [lines (a-c) of Fig. 2S] and $\hbar\gamma/J = 20$ [line (d) of Fig. 2S]. Changing $N$ from $N = 40$ to $N = 20$ merely modifies the shape of the single-particle correlations, but not their decay rate and periodicity. We also see that the period is equal to $2\pi\hbar/U$, so that doubling $U$ decreases the period of the oscillations by two [see for example line (b) for $U/J = 10$ and line (a) for $U/J = 20$].

In the following we describe how the single-particle cor-

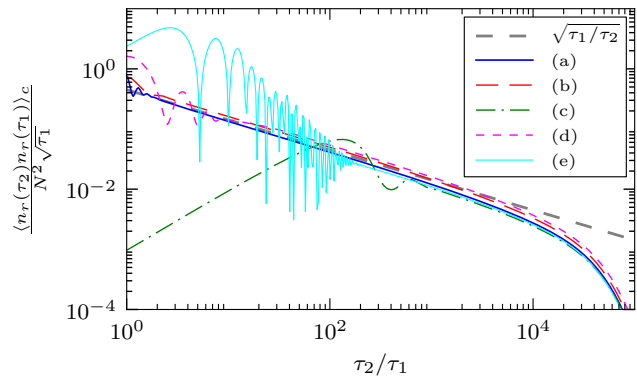

Figure 1S. (color online) Rescaled two-time density-density correlations versus ratio of rescaled times at fixed $\tau_1 = 1.5 \times 10^{-6}$. Parameters $(U/J, \hbar\gamma/J, N) = (20, 5, 50), (20, 20, 50), (1, 5, 50), (20, 5, 20), (1, 0.1, 50)$ for (a), (b), (c), (d), (e), respectively. The dashed line indicates the scaling behavior in $(\tau_2/\tau_1)^{-1/2}$ expected in the aging regime.

relations $\langle b_r^\dagger(t_2) b_r(t_1)\rangle = \text{Tr}\left[b_r^\dagger e^{\mathcal{L}_0^{\Omega_b}(t_2-t_1)} \left(b_r e^{\tilde{\mathcal{L}}^{\Lambda_0} t_1} \rho_0\right)\right]$ can be approximately determined within the adiabatic elimination method. The first evolution until time $t_1$, given by $e^{\tilde{\mathcal{L}}^{\Lambda_0} t_1} \rho_0$, is computed in the decoherence free subspace using the effective evolution equation (4). After the application of $b_r$, the density matrix belongs to the subspace $\Omega_b$, where it evolves between times $t_1$ and $t_2$. The elements of the subspace $\Omega_b$ can be represented by $\rho_b^\Omega = \sum_{n_l} \left(\rho_{n_l}^{(0)} |n_l, n_r - 1\rangle\langle n_l, n_r| + \rho_{n_l-1}^{(1)} |n_l - 1, n_r\rangle\langle n_l, n_r|\right)$ with $n_r = N - n_l$.

If we neglect all transitions to subspaces with higher dissipative rates, the evolution within the subspace $\Omega_b$ is given by the coupled equations:

$$\frac{d}{dt}\rho_{n_l}^{(e)} = \left\{-\frac{\gamma}{2} + i\frac{U}{\hbar}\left[\delta_{e,0}(N - 2n_l - 1) + n_l\right]\right\}\rho_{n_l}^{(e)}$$
$$+ i\frac{J}{\hbar}\sqrt{n_l + e}\sqrt{n_r - e}\rho_{n_l+2e-1}^{(1-e)} - i\frac{J}{\hbar}\sqrt{n_l + 1}\sqrt{n_r}\rho_{n_l}^{(1-e)}$$
(8S)

These results contain the main features seen in Fig. 2S (not shown).

## DYNAMICS OF A DISSIPATIVELY PREPARED BCS DARK STATE

In this section, we derive the results for the Green function and anomalous Green function for a s-wave BCS state, both for the unitary evolution and for a dissipative evolution.

*Unitary Evolution* – In diagonal form, the BCS Hamiltonian is $H_{\text{BCS}} = \sum_{\vec{k},\sigma} E_{\vec{k}} \gamma_{\vec{k},\sigma}^\dagger \gamma_{\vec{k},\sigma} + const.$ [9]. Here

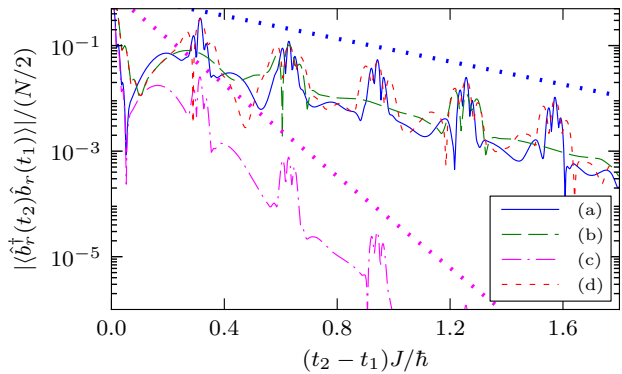

Figure 2S. (color online) Evolution of the single particle correlations as a function of $t_2 - t_1$ from the master equation, Eq. (2) of the main text, for $\tau_1 = 10^{-3}$ ($t_1$ is fixed in rescaled times). Parameters $(U/J, \hbar\gamma/J, N) = (20, 5, 40), (10, 5, 40), (20, 20, 40), (20, 5, 20)$ for (a), (b), (c), (d) respectively. The correlations are dominated by an exponential decay, the function $e^{-\gamma(t_2-t_1)/2}$ is indicated by thick dotted lines for $\hbar\gamma/J = 5, 20$. There are underlying coherent oscillations of periodicity $2\pi\hbar/U$ caused by the interaction.

the Bogoliubov operators $\gamma_{\vec{k},\sigma}$ are defined by $c_{\vec{k},\uparrow} = u_{\vec{k}}^* \gamma_{\vec{k},0} + v_{\vec{k}} \gamma_{\vec{k},1}^\dagger$ and $c_{-\vec{k},\downarrow}^\dagger = -v_{\vec{k}}^* \gamma_{\vec{k},0} + u_{\vec{k}} \gamma_{\vec{k},1}^\dagger$, where $|u_{\vec{k}}|^2 + |v_{\vec{k}}|^2 = 1$ and $c_{\vec{k},\sigma}$ annihilates a fermion with momentum $\vec{k}$ and spin $\sigma$. For the ground state of the s-wave BCS Hamiltonian, the coefficients $u_{\vec{k}}$ and $v_{\vec{k}}$ can be chosen to be real. In our conventions they obey $v_{\vec{k}}^2 = 1 - u_{\vec{k}}^2 = \frac{1}{2}\left(1 - \frac{\epsilon_{\vec{k}} - \mu}{E_{\vec{k}}}\right)$. The energy is $E_{\vec{k}} = \sqrt{\Delta^2 + (\epsilon_{\vec{k}} - \mu)^2}$, with the non-interacting electron dispersion relation $\epsilon_{\vec{k}}$, the chemical potential $\mu$, and the gap $\Delta$, which in mean-field theory has to be determined self-consistently. The ground state of the Hamiltonian is $|\text{BCS}\rangle = \prod_{\vec{k}} \left(u_{\vec{k}} + v_{\vec{k}} c_{\vec{k},\uparrow}^\dagger c_{-\vec{k},\downarrow}^\dagger\right)|0\rangle$, where $|0\rangle$ is the vacuum for the $c_{\vec{k},\sigma}$ operators. The low energy single particle excitations are the Bogoliubov excitations, created by the $\gamma_{\vec{k},\sigma}^\dagger$ operators. The Green function for the BCS state is given by $G_\uparrow^>(\vec{k}, t_2, t_1) = |u_{\vec{k}}|^2 e^{-iE_{\vec{k}}(t_2-t_1)/\hbar}$ and the anomalous Green function is $G^A(\vec{k}, t_2, t_1) = -u_{\vec{k}}^* v_{\vec{k}} e^{-iE_{\vec{k}}(t_2-t_1)/\hbar}$. Here it is apparent that, under the unitary evolution, only a time-dependent phase is acquired, whereas the absolute values of the Green functions remain constant.

*Dissipative evolution* – Let us now turn to a BCS state which is engineered as the dark state of a dissipative dynamics. For simplicity we consider a solely dissipative evolution determined by $\mathcal{D}$ while the Hamiltonian vanishes, $H = 0$. The quantum jump operators are chosen to be the Bogoliubov annihilation operators $K_{\vec{k},\sigma} = \gamma_{\vec{k},\sigma}$. The master equation for the evolution of a density matrix is then given by

$$\frac{\partial}{\partial t}\rho = \mathcal{D}(\rho) = \frac{\kappa}{2} \sum_{\vec{k},\sigma} \left(2\gamma_{\vec{k},\sigma} \rho \gamma_{\vec{k},\sigma}^\dagger - \gamma_{\vec{k},\sigma}^\dagger \gamma_{\vec{k},\sigma} \rho - \rho \gamma_{\vec{k},\sigma}^\dagger \gamma_{\vec{k},\sigma}\right) \tag{9S}$$

where $\kappa$ is a coupling constant. The BCS state, being the vacuum state of the Bogoliubov operators ($\gamma_{\vec{k},\sigma}|\text{BCS}\rangle = 0$ for all $\vec{k}$ and $\sigma$), is a dark state of the chosen dissipative coupling and does not evolve in time.

The Green function in this dark state then reads

$$G_\uparrow^>(\vec{k}, t_2, t_1) = \text{Tr}\left[c_{\vec{k},\uparrow} e^{\mathcal{D}(t_2-t_1)}\left(c_{\vec{k},\uparrow}^\dagger |\text{BCS}\rangle\langle\text{BCS}|\right)\right]$$
$$= |u_{\vec{k}}|^2 e^{-\frac{\kappa}{2}(t_2-t_1)} \tag{10S}$$

As stated in the main text, we obtain that the Green function (10S) decays exponentially. Using a similar argument, one can see that the anomalous Green function also decays exponentially $G^A(\vec{k}, t_2, t_1) = -u_{\vec{k}}^* v_{\vec{k}} e^{-\frac{\kappa}{2}(t_2-t_1)}$.



\end{document}